\documentclass[twocolumn, preprintnumbers,superscriptaddress,prb]{revtex4}
\usepackage[latin1]{inputenc}
\usepackage[american]{babel}
\usepackage{graphicx}                                                   
\usepackage{dcolumn}
\usepackage{bm}

\begin{document}                

\author{Alexander Hinderhofer} \affiliation{Institut für Angewandte Physik
  Universit\"at T\"ubingen, Auf der Morgenstelle 10, 72076 T\"ubingen,
  Germany}

\author{Ute Heinemeyer} \affiliation{Institut für Angewandte Physik
  Universit\"at T\"ubingen, Auf der Morgenstelle 10, 72076 T\"ubingen,
  Germany}

\author{Alexander Gerlach} \affiliation{Institut für Angewandte Physik
  Universit\"at T\"ubingen, Auf der Morgenstelle 10, 72076 T\"ubingen,
  Germany}

\author{Stefan Kowarik} \affiliation{Institut für Angewandte Physik
  Universit\"at T\"ubingen, Auf der Morgenstelle 10, 72076 T\"ubingen,
  Germany}

\author{Robert M. J. Jacobs} \affiliation{Chemistry Research Laboratory,
  University of Oxford, Oxford OX1 3TA, U.K.}

\author{Youichi Sakamoto} \affiliation{Institute for Molecular Science
  Myodaiji, Okazaki 444-8787, Japan}

\author{Toshiyasu Suzuki} \affiliation{Institute for Molecular Science
  Myodaiji, Okazaki 444-8787, Japan}

\author{Frank Schreiber}\email{frank.schreiber@uni-tuebingen.de}
\affiliation{Institut für Angewandte Physik Universit\"at T\"ubingen, Auf der
  Morgenstelle 10, 72076 T\"ubingen, Germany}

\date{\today}
\title{Optical Properties of Pentacene and Perfluoropentacene Thin Films}

\begin{abstract}
  The optical properties of pentacene (PEN) and perfluoropentacene(PFP) thin
  films on various SiO$_{2}$ substrates were studied using variable angle
  spectroscopic ellipsometry. Structural characterization was performed using
  X-ray reflectivity and atomic force microscopy. A uniaxial model with the
  optic axis normal to the sample surface was used to analyze the ellipsometry
  data. A Strong optical anisotropy was observed and enabled the direction of
  the transition dipole of the absorption bands to be determined. Furthermore,
  comparison of the optical constants of PEN and PFP thin films with the
  absorption spectra of the monomers in solution shows significant changes due
  to the crystalline environment. Relative to the monomer spectrum the
  HOMO-LUMO transition observed in PEN (PFP) thin film is reduced by 210~meV
  (280~meV). Surprisingly, a second absorption band in the PFP thin film shows
  a slight blueshift~(40~meV) compared to the spectrum of the monomer with its
  transition dipole perpendicular to that of the first absorption band.

\end{abstract}

\maketitle


\section{Introduction}
One of the reasons for the substantial application potential of organic
semiconducting molecules is their tunability, by exchanging certain functional
groups of a molecule while leaving the backbone unchanged.  For organic
electronics in general and organic field effect transistors in particular
pentacene (C$_{22}$H$_{14}$, PEN) is the most popular compound~\cite{brutting,
  scholz}, although there are many materials and compounds to choose from, and
it is not entirely obvious why PEN would actually have to be the best choice.
Possible alternatives to PEN may be, e.g., rubrene, which has its own
structural subtleties~\cite{Kafer, KowRUB, science} and diindenoperylene,
which exhibits excellent structural order~\cite{KowDIP, Durr, Pflaum, karl}.

Another option is to stay with the PEN backbone and study modifications of
PEN. Recently, perfluorinated pentacene (C$_{22}$F$_{14}$, PFP) has been
identified as an interesting, possibly complementary option to
PEN~\cite{Sakamoto_2004_JotACS_126_8138, Sakamoto_2006_MCLC_444_225,
  Inoue_2005_JJoAP_44_3663, Koch_2007_AM_19_112}.  Due to the strong
electronegativity of the fluorine atoms the charge transport behavior changes
from p-type into n-type, which opens the possibility of low-stress bipolar
transistors. While there have been pioneering studies on PFP, its properties
still require thorough investigation.

Even for PEN, despite the efforts in recent years, several issues regarding
the growth, structure, and phase behavior are still under
investigation~\cite{Ruiz_2004_CM_16_4497,KowarikTSF,Bouchoms_1999_SM_104_175,Mayer_2006_PRL_97_105503}.
Thin films of pentacene on
SiO$_{2}$~\cite{Bouchoms_1999_SM_104_175,Mayer_2006_PRL_97_105503} and also on
glass exhibit a coexistence of the 'thin film phase' and the 'bulk phase', the
latter of which could be identified by a smaller $d(001)$ spacing. The
substrate and its surface treatment influence the grain size and morphology,
for example on glass the grain size is reported to be smaller than on oxidised
silicon~\cite{Puigdollers_2003_TSF_427_367,Kim_2005_CAP_6_925}.

Structural issues significantly impact the charge carrier mobility, which is
the crucial quantity determining the frequency at which organic semiconductor
devices can be operated. The optical properties in the condensed phase can, of
course, also be strongly affected by the film structure because of the
differences in the coupling of a given molecule with its environment,
particularly for crystalline films of PEN and PFP. Due to the intimate
relationship between the optical spectrum and electronic properties, efforts
to improve our understanding of the material and the device performance have
to include the optical properties, which in crystalline materials are
generally anisotropic.

Recently, two studies focused on the optical properties of rather thick PEN
films (100 and 44~nm)~\cite{Faltermeier_2006_PRB_74_125416,
  Park_2002_APL_80_2872} using isotropic models. The role of possible
anisotropies as well as the behavior of thinner films and other growth regimes
of PEN require further investigation. Specifically because thinner films
exhibit a significant lower amount of the 'bulk phase' and are dominated by
the 'thin film phase', these issues remain open.

\begin{figure} [ht]
        \begin{center}
        \includegraphics [width=8.5cm] {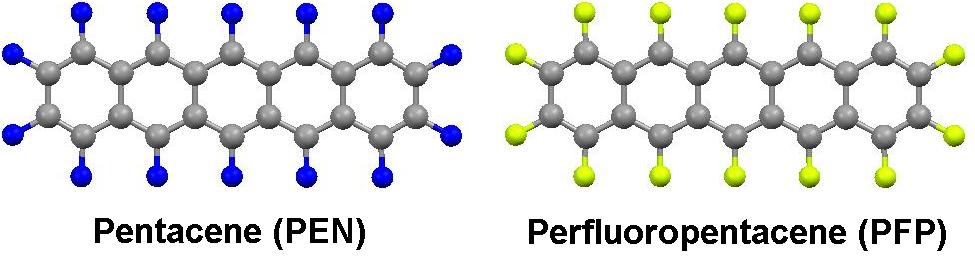}
        \caption{Scheme of the PEN~(C$_{22}$H$_{14}$) and PFP~(C$_{22}$F$_{14}$) molecule.}
        \label{mol}
        \end{center}
\end{figure}

This paper is devoted to a comparison of the optical properties of PFP and PEN
(Fig.~\ref{mol}) thin films grown by organic molecular beam deposition
(OMBD)~\cite{Schreiber04Orga,Witte04Grow}.  This should serve, first, to
establish a more solid data base for PEN and, second, to shed light on the
properties of PFP as a rather new and promising material for organic
electronics and optoelectronics.

\section{Experimental}

Three different types of substrates were used, a $\sim$1~mm thick Si(100)
wafer with a native oxide layer ($d \sim$1~nm), a silicon wafer with a thermal
SiO$_{2}$ layer ($d=430$~nm) and a glass slide ($d=0.5$~mm). All substrates
were cleaned with acetone and isopropanol in an ultrasonic bath and then
rinsed with ultra pure water. The silicon oxide thickness (in the case of the
Si wafers) and optical constants of the substrates were determined by
ellipsometry prior to film growth.  Organic molecular beam deposition (OMBD)
under ultra high vacuum conditions (base pressure $2 \times 10^{-10}$~mbar)
was used to grow PEN and PFP thin films by thermal evaporation from a Knudsen
cell. The growth rate of about 2~\AA /min was monitored via a water-cooled
quartz crystal microbalance and the substrate temperature was kept constant at
$T=30\, ^\circ$C. Under these growth conditions PEN is structurally well
ordered~\cite{Ruiz_2004_CM_16_4497}.

The ellipsometry~\cite{azzam} data were taken \textit{ex-situ} in air, within
a few hours after growth, with a Woollam ellipsometer (M-2000, rotating
compensator type) in an energy range from $1.25-3.5$~eV. The spectra were
recorded by a CCD camera with a wavelength resolution of 1.6~nm.  Two
different types of measurements were performed, namely reflection and
transmission ellipsometry. For reflection ellipsometry an automated goniometer
was used to record data at 13 different angles of incidences between
$40^{\circ}$ and $80^{\circ}$.  In transmission we measured at five different
angles between $20^{\circ}$ and $70^{\circ}$, manually aligned with an error
of $\pm 0.5^{\circ} $.  The data were analyzed based on established routines
using a commercial software (WVASE32)~\cite{Woo,Kytka}.  \textit{In-situ}
X-ray reflectivity measurements ($\lambda = 0.92$\AA) were performed similar
to Ref.~\onlinecite{KowDIP} in UHV at beamline ID10B of the ESRF in Grenoble,
France. Details will be published in Ref.~\onlinecite{Kowarik_unpub}.

Absorption spectra of PEN and PFP were recorded with a Varian Cary50 UV-VIS
spectrometer. For this PEN and PFP material were dissolved at low
concentration in dichlorobenzene and measured immediately in order to avoid
dissociation effects which could be observed after several hours.  Tapping
mode AFM-measurements were performed in air about two to four months after
deposition using a Digital Instruments Multimode AFM.


\section{Results}

\subsection{Structure and Morphology}
\begin{figure} [ht]
        \begin{center}
        \includegraphics [width=8.5cm] {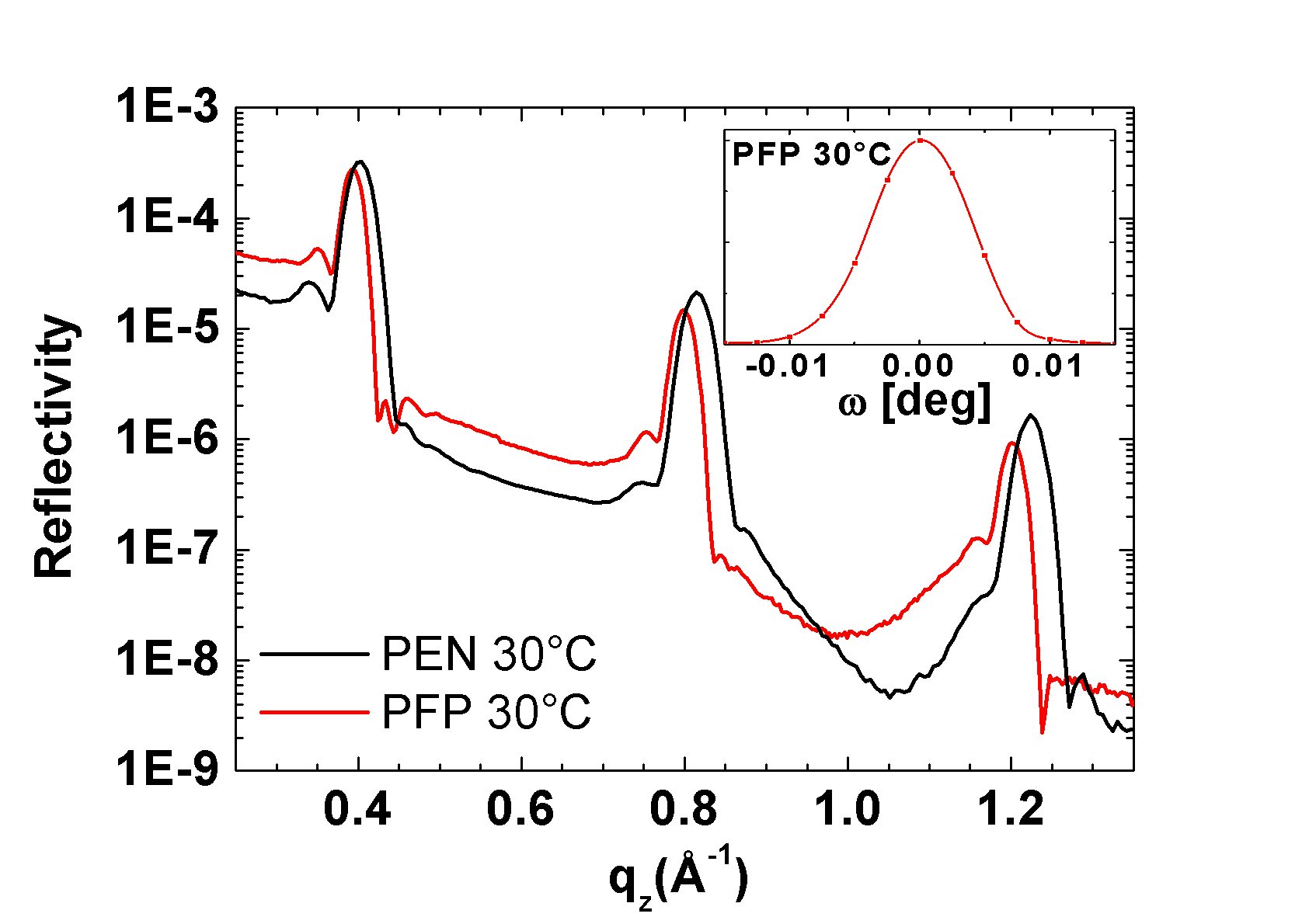}
        \caption{X-ray reflectivity data for PEN ($d =16$~nm) and PFP ($d = 25$~nm) grown at $30\, ^{\circ}$C substrate temperature on a silicon wafer with native oxide. The inset shows the narrow rocking width at the first Bragg reflection of PFP.}
        \label{spec}
        \end{center}
\end{figure}

As has been demonstrated before PEN~\cite{KowarikTSF} and PFP grow in
crystalline films. The film structure was studied by X-ray reflectivity and
AFM, as complementary techniques, with the purpose of supporting and
complementing the optical spectra obtained by spectroscopic ellipsometry.
Out-of-plane X-ray measurements of the first three Bragg reflections of PEN
and PFP are shown in Fig.~\ref{spec}.  The pronounced Bragg reflections with
Laue oscillations and the width of the rocking curve at the first Bragg
reflection of PFP ($0.0089 ^\circ$) demonstrate the high structural order of
the thin films. From the position of the second and third Bragg reflections
the out-of-plane $d(001)$ lattice spacing is determined to be 15.4~\AA~in PEN
and 15.7~\AA~in PFP, in agreement with the PFP thin film X-ray data from
Ref.~\onlinecite{Inoue_2005_JJoAP_44_3663}.

Both for PFP and PEN in the specular reflectivity in Fig.~\ref{spec} Bragg
reflections corresponding to only one polymorph are observed.  For PEN though
grazing incidence X-ray diffraction (GIXD) reveals small traces of a second
phase. Apart from this issue of phase coexistence in PEN, the structural motif
appears to be rather similar for PFP and PEN for growth at $T=30\, ^{\circ}$C.
\begin{figure} [ht]
        \begin{center}
        \includegraphics [width=8.5cm] {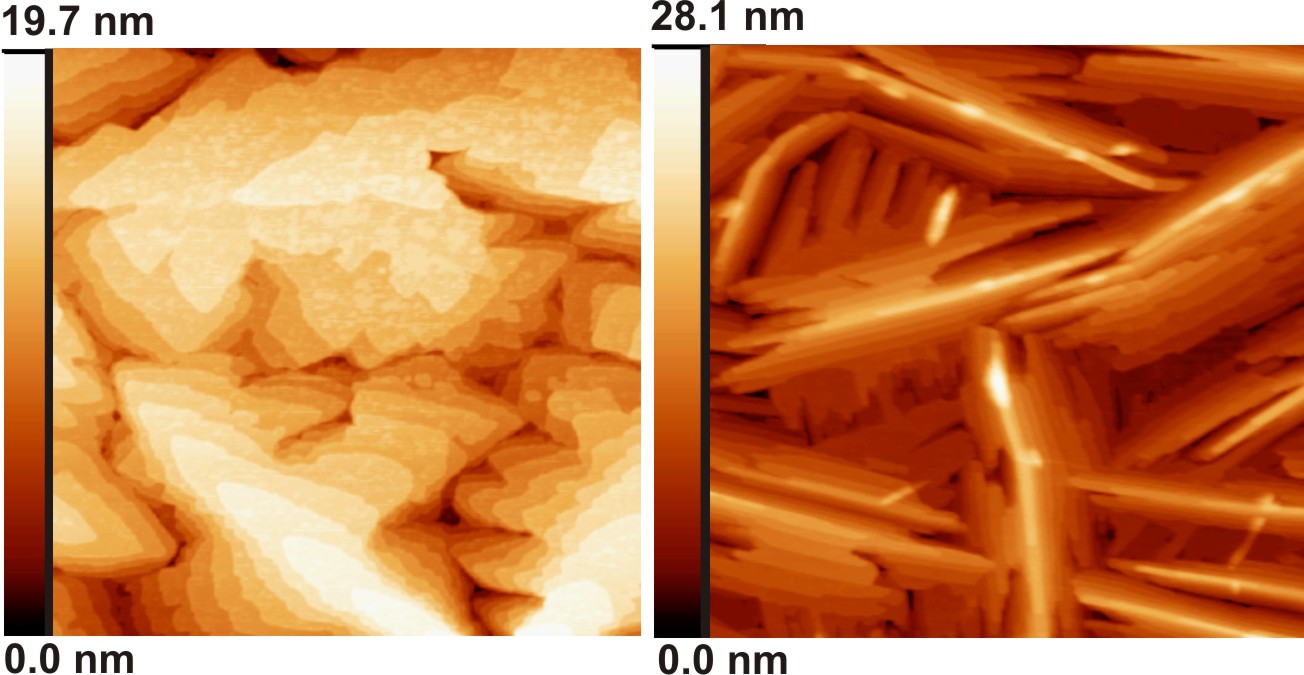}
        \caption{Typical morphologies for PEN and PFP thin films grown at $30\, ^{\circ}$C on native SiO$_{2}$ at 22.5~nm thickness. Each picture shows an area of 3$\,\mu$m$\, \times \,$3$\, \mu$m. Left: Large grains of PEN. Right: PFP grows in lamellar shaped grains.}
        \label{afm}
        \end{center}
\end{figure}    

AFM data from PEN and PFP (Fig.~\ref{afm}) show the morphology of the thin
films on native SiO$_{2}$. Both thin films consist of randomly oriented
crystalline grains. Steps between monolayers are observable with a typical
height of $1.6 - 2.2$~nm for PEN and $1.6 - 1.9$~nm for PFP. These almost
correspond to the $d(001)$ spacing as determined from X-ray measurements.
Obviously, on isotropic substrates such as oxidized silicon, there is no
azimuthal order, which results in an in-plane isotropic film structure.

\subsection{Optical properties}
Due to the high out-of-plane order and the in-plane isotropy both systems
exhibit uniaxial optical properties with the optic axis perpendicular to the
substrate surface. The determination of the optical constants for this
geometry is a widely discussed issue in ellipsometry, see e.g.\ 
Ref.~\onlinecite{Campoy-Quiles_2005_PRB_72_045209}. Generally, an isotropic
fit approach will not give the average properties of both
axes~\cite{Smet_1994_JoAP_76_2571} but will produce artificial absorption
features~\cite{Dignam_1971_TotFS_67_3306}.

For the samples presented here (silicon substrate with native oxide and a
$\sim $20~nm thin uniaxial film on top) standard Variable Angle Spectroscopic
Ellipsometry (VASE) is not sensitive enough for determining a unique set of
four optical constants, i.e.\ real and imaginary part of the in-plane and
out-of-plane component at each
wavelength~\cite{Smet_1994_JoAP_76_2571,Bortchagovsky_1997_TSF_307_192}.
Therefore we used two techniques to increase the sensitivity to the anisotropy
of our samples. In both cases we combined data from a thin film grown on
native oxide with data from a simultaneously grown film on another substrate
to perform a multisample analysis.  First, using a substrate with a thick
thermal oxide (430~nm) increases the sensitivity, because the intermediate
layer decorrelates data measured at different angles of
incidence~\cite{Bortchagovsky_1997_TSF_307_192}. The drawback of this method
is that the exact determination of the thermal oxide optical constants is
crucial for the analysis of organic thin films. Therefore it is useful to
compare the results with an independent second approach where reflection data
are combined with complementary transmission ellipsometry data from a glass
sample~\cite{Ramsdale_2002_AM_14_212}. In this case, the thickness of the
'roughness layer' within the ellipsometric model must be adjusted accounting
for the higher film roughness on glass.

Our layer model for the ellipsometry data is from bottom to top: substrate
([silicon + oxide] or glass) // uniaxial thin film // Effective Medium
Approximation EMA (mixing void and uniaxial thin film 50\%). The thicknesses
of the EMA layers are defined by AFM-measurements, i.e. 8.5~nm for PEN and
10~nm for PFP for the samples studied. The AFM images show that the thin films
have structures on the length scale of the wavelength of light used for
ellipsometry, for this case an EMA model usually gives only an approximate
representation of the morphology. The effective thicknesses of the thin films
are determined isotropically in the transparent range by parameterization of
the refractive index dispersion with a Cauchy function (22.5~nm). The uniaxial
results obtained with both multisample approaches agree with each other, both
for PEN as well as for PFP thin films. Thus we are able to determine the
direction of the transition dipole $\vec{P} = (p_{\perp},p_{\parallel})$ for
each band.

We do not observe significant differences between ellipsometry data measured
\textit{in-situ} under UHV conditions and ellipsometry data measured
\textit{ex-situ} in air, hence we can exclude strong oxidation effects as they
occur for instance in rubrene~\cite{Kytka} .

                                
\subsubsection{Comparison of PFP and PEN thin films.} 
\begin{figure} [ht]
                \begin{center}
                \includegraphics [width=8.5cm] {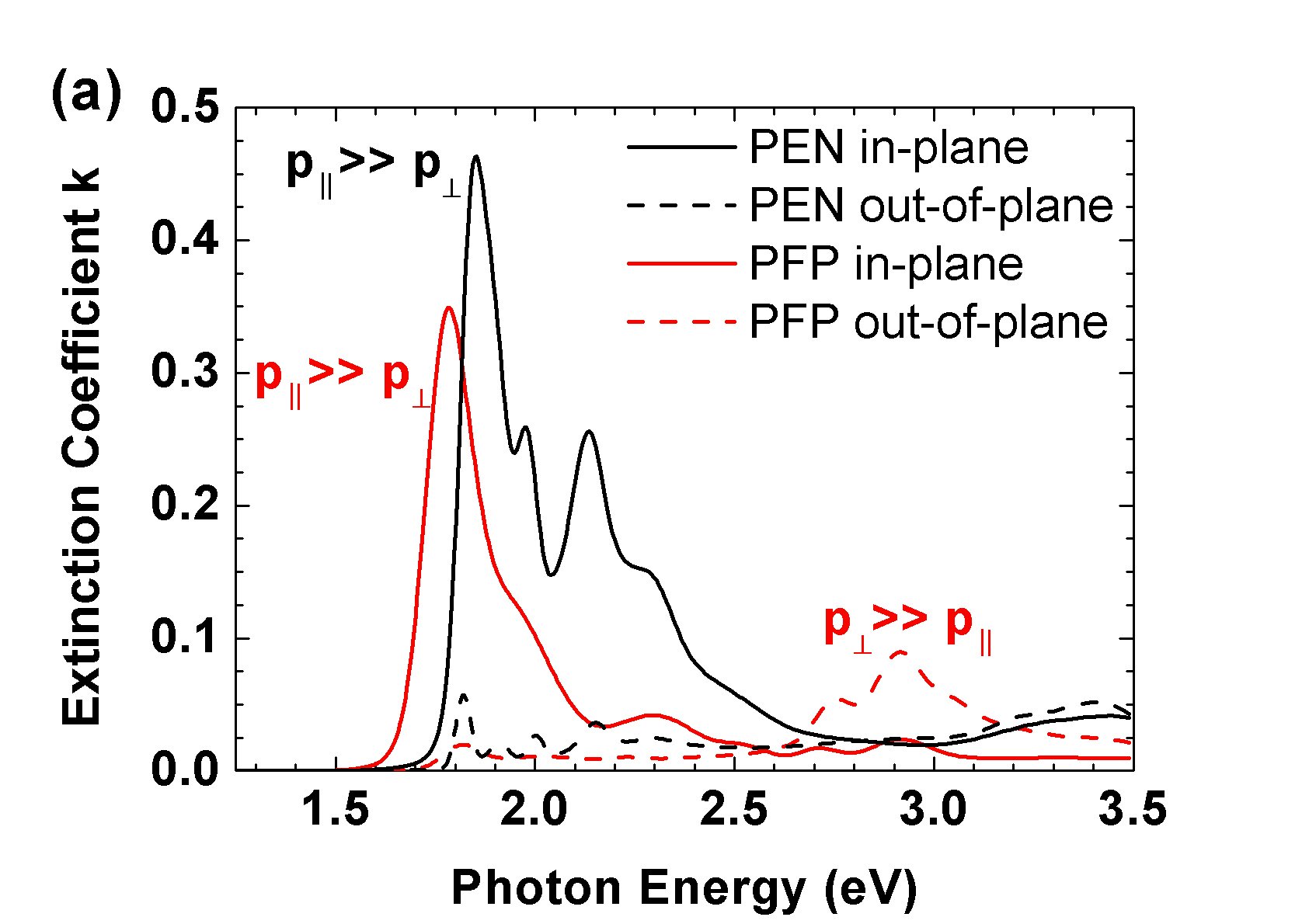}
                \includegraphics [width=8.5cm] {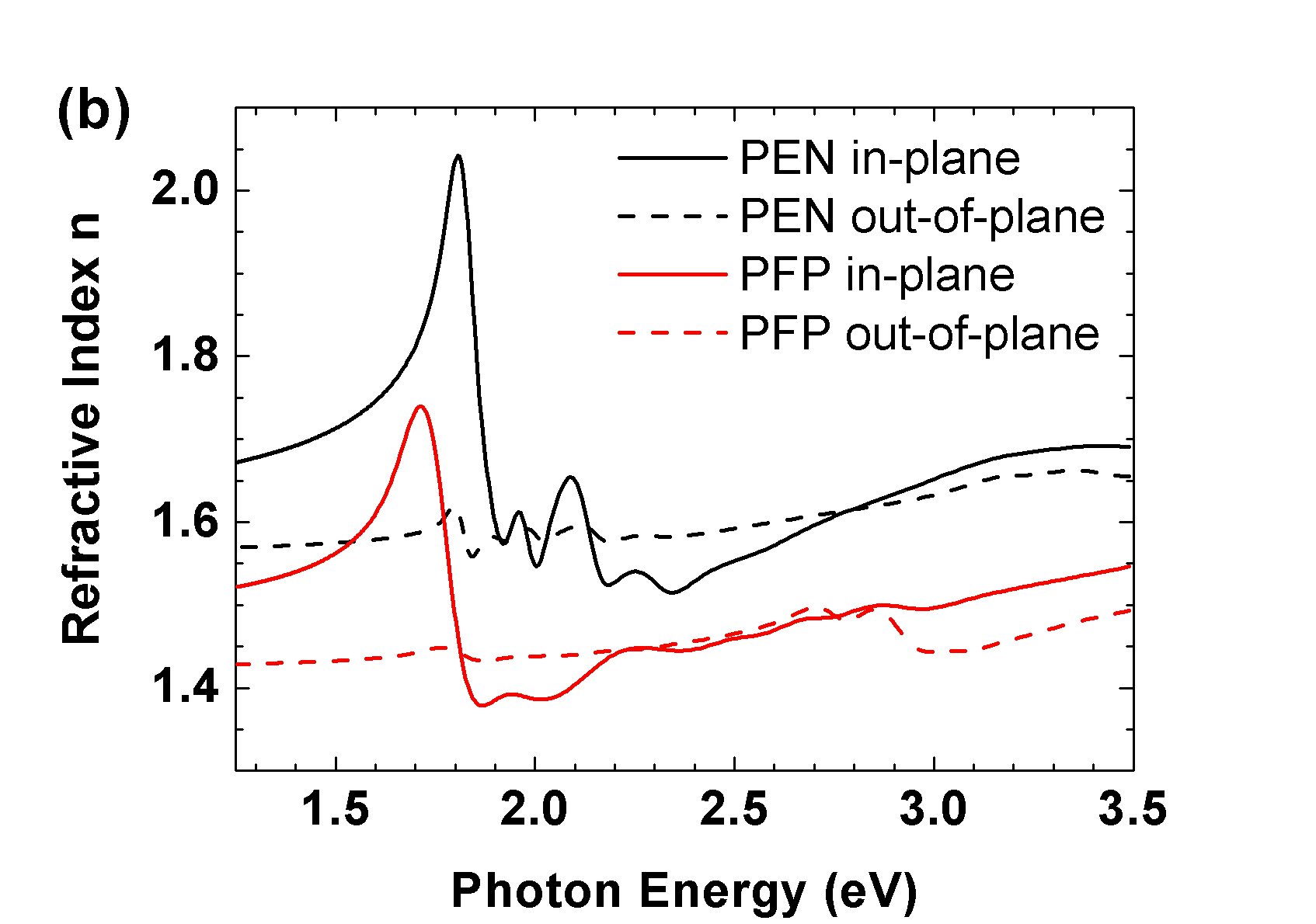}
                \caption{Optical constants for the in-plane and out-of-plane
                components of 22.5~nm PEN and 22.5~nm PFP thin films grown at
                $30\, ^{\circ}$C substrate temperature, obtained by
                multisample analysis. (a)~Extinction
                coefficients. (b)~Refractive indices.}
                \label{ext}
                \end{center}
\end{figure}

Fig.~\ref{ext} displays the spectra as obtained with the above multisample
approach for PEN and PFP thin films, respectively.  In the energy range of
$1.5-2.6$~eV we concentrate in the following on the in-plane component, since
this is the dominating contribution and under the given experimental
conditions the orthogonal out-of-plane component is very weak. The latter is
at least five times smaller than the in-plane component and thus difficult to
resolve reliably (relative error 50$\%$).
In this energy range the out-of-plane component is probably a small projection
of the absorption band with the transition dipole along the short axis of the
PEN and PFP molecules on the surface normal ($p_{\|}\gg p_{\perp} $). This is
also in good agreement with the tilt angle of pentacene molecules on
SiO$_{2}$, i.e.\ $10 \pm 5^{\circ}$ with respect to the surface normal, and
with the idea that the transition dipole is essentially perpendicular to the
long axis of the molecule~\cite{Yoshikawa_2006_SS_600_2518}. At energies above
2.6~eV the out-of-plane component exceeds the in-plane component in case of
PFP and is thus well defined.

The error for the extinction coefficient $k$ peak positions is $\pm 0.01$~eV,
the error for the absolute values of $k$ is $\pm 0.05$ for both PEN and PFP.
The absolute error for $n_{\infty}$ is approximately $\pm 0.05$ for all four
determined refractive indices. The refractive index shown in Fig.~\ref{ext} is
Kramers-Kronig consistent with $k$.

In Fig.~\ref{ext} the PEN spectrum shows a pronounced first peak whose energy
position of 1.85~eV indicates the optical band gap in agreement with
absorption measurements on glass and previous ellipsometry
measurements~\cite{ostro, Faltermeier_2006_PRB_74_125416} The spectral feature
at 1.97~eV has been assigned to
Davydov-splitting~\cite{Hesse_1980_CP_49_201,Lee_1977_CPL_51_120, ostro} but
is currently still debated~\cite{Grobosch_2006_PRB_74_155202}. The subsequent
peaks above 2~eV exhibit an equal energy spacing of $170\pm10$~meV and are
therefore likely to be associated with a vibronic progression.

The effect of substituting all hydrogen atoms by fluorine atoms results in
interesting changes of the extinction coefficient.  The optical band gap is
red-shifted by 70~meV in PFP compared to PEN. While the shift of the second
peak ($\sim 1.94$~eV) in PFP with only $\sim 30$~meV is smaller, it is much
broader and appears therefore as a shoulder of the first peak. Interestingly,
the third peak of the PEN spectrum is not visible in the PFP spectrum and only
the fourth (2.30~eV) and fifth ($2.48$~eV) feature can be observed with
similar energy spacing and position as in PEN. Another significant difference
between PEN and PFP thin film spectra can be observed at energies above
2.6~eV. While for PEN there is no further absorption band visible in either
component, new features appear for PFP. They can be found in both components,
but the extinction coefficient is about four times bigger in the out-of-plane
component showing that the transition dipole lies approximately perpendicular
to the sample surface ($p_{\perp}\gg p_{\|}$), i.e.\ along the long axis of
the molecule, in contrast to the absorption band at lower energies.
                  

\subsubsection{Solvent spectra of PEN and PFP}
\begin{figure} [ht]
        \begin{center}
        \includegraphics [width=8.5cm] {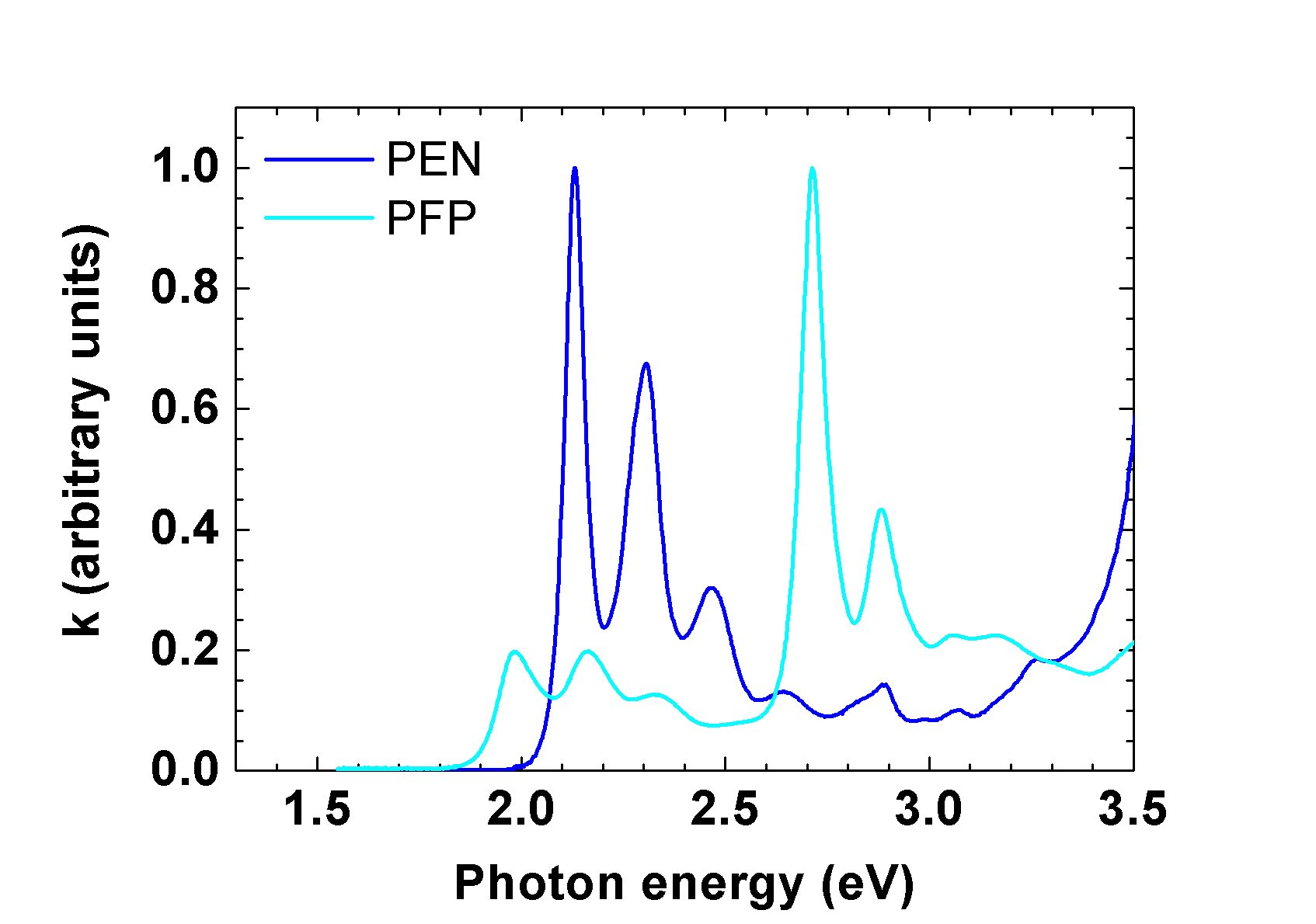}
        \caption{Normalized absorption spectra of PEN and PFP dissolved in dichlorobenzene.}
        \label{sol}
        \end{center}
\end{figure}

Fig.~\ref{sol} shows the absorption spectra of PEN and PFP in dichlorobenzene.
Since the absolute values are not obtained by this measurement, the spectra
for PFP and PEN are normalized to detect changes in relative amplitudes
between both molecules. Our measurements are in agreement with
Ref.~\onlinecite{Sakamoto_2004_JotACS_126_8138}.  In the PFP solvent spectra
two absorption bands with a vibronic progression (energy spacing
$170\pm10$~meV) are observed. The second band starting at 2.71~eV (assigned to
the $S_0\rightarrow S_3$ transition in Ref.~\onlinecite{Medina_2007}) is more
intense than the first at 1.99~eV ($S_0 \rightarrow S_1$). In contrast, the
$S_0\rightarrow S_1$ absorption band of PEN at 2.13~eV is much more intense
than the $S_0 \rightarrow S_3$ band at $\sim 2.85$~eV.  The energy spacing of
the first absorption band is similar to that of PFP, which implies that the
vibronic progression for both molecules depends mainly on the vibrations of
the carbon core. This is reasonable since the lengths of the C-C bonds change
significantly compared to the lengths of the C-H bonds for a HOMO-LUMO
transition of PEN~\cite{scholz}, i.e.\ the HOMO-LUMO transition couples more
strongly to the carbon core.  Obviously, the fluorination changes the
resulting oscillator strengths of the absorption bands substantially, which is
reproduced by time-dependent density functional theory
calculations~\cite{Medina_2007}.

\subsubsection {Comparison of thin films and solution spectra}                  
\begin{figure} [ht]
        \begin{center}
        \includegraphics [width=8.5cm] {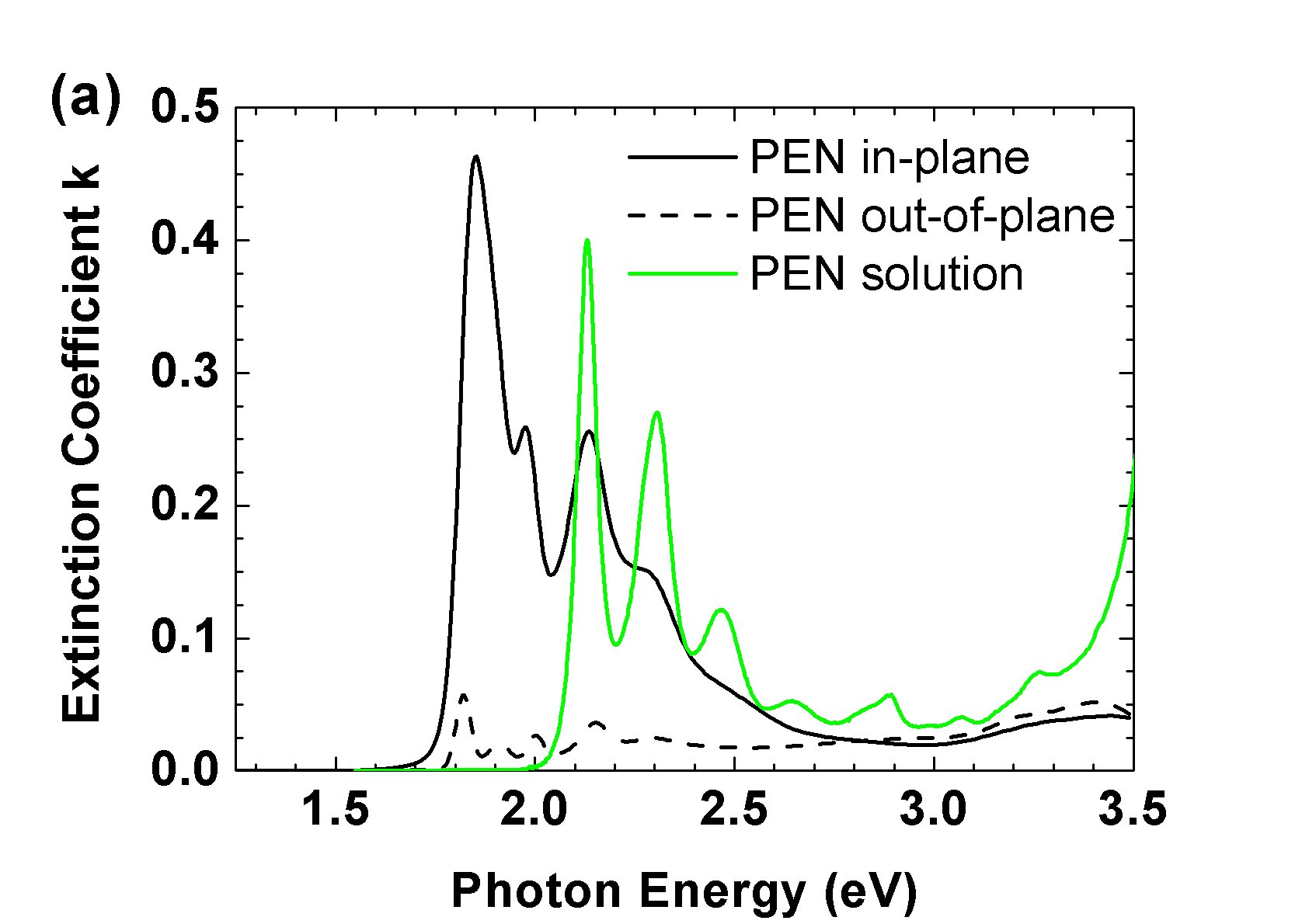}
        \includegraphics [width=8.5cm] {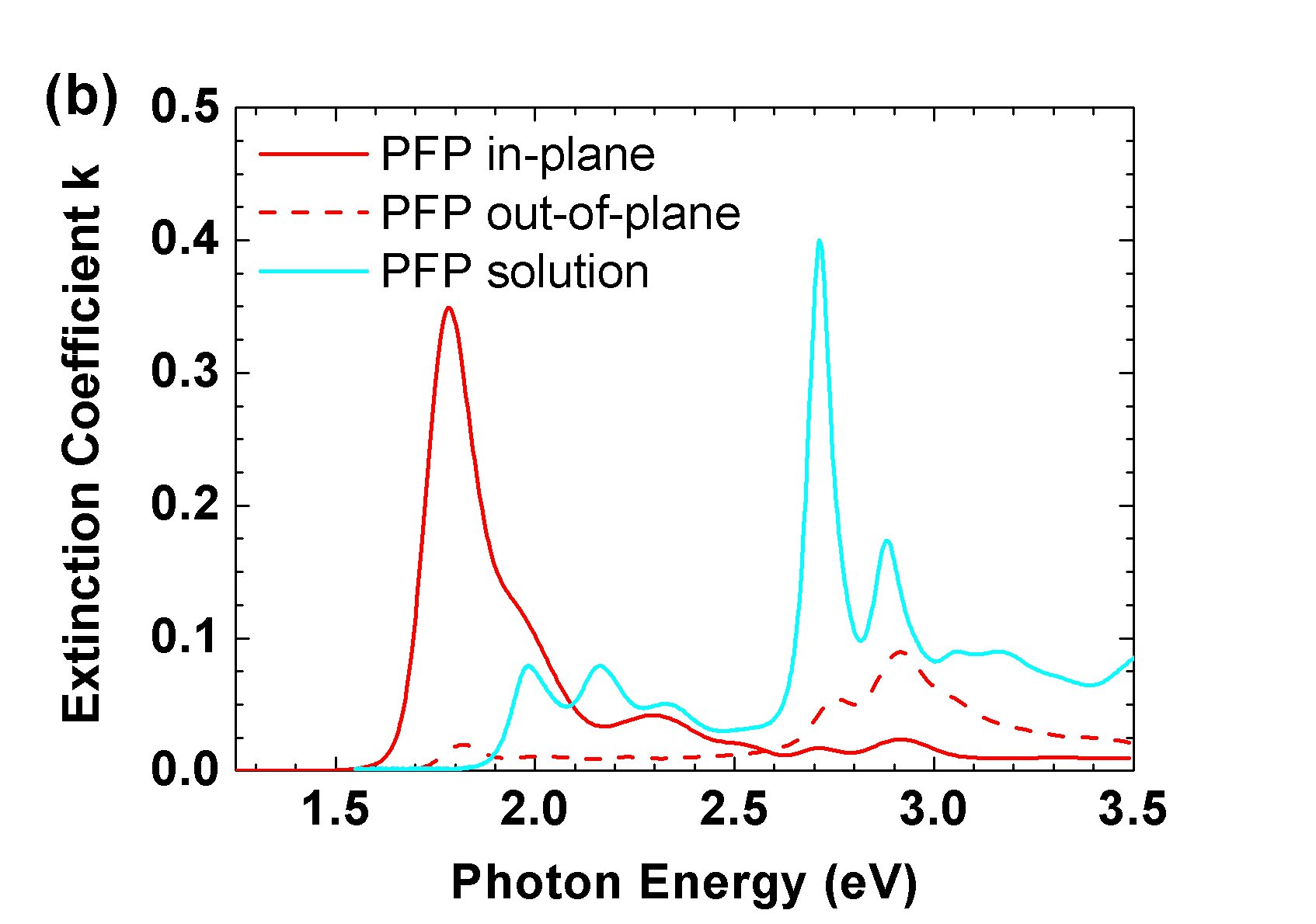}
        \caption{ Thin film spectra of 22.5~nm thick films, growth temperature
          $30\, ^{\circ}$C and spectra in solution. The intensity for the
          spectrum in solution is given in arbitrary units. a) PEN b) PFP:
          Note that around 3~eV the out-of-plane component can be determined
          reliably, since it is actually rather strong in this energy range.}
        \label{PEN}
        \end{center}
\end{figure}

The comparison of the crystalline thin film spectra to the spectra of the
monomer in solution gives information about the changes of the spectra induced
by the crystal structure and the coupling of the molecules. This is shown in
Fig.~\ref{PEN}. For both molecules in solution one can clearly distinguish the
vibronic transitions, whereas in the thin film all features are smeared out.
For both materials the HOMO-LUMO absorption band in thin films is shifted to
lower energies with respect to the monomer spectrum, i.e.\ 210~meV in case of
PFP, and 280~meV in case of PEN. The second absorption band shifts slightly to
the blue (40~meV) for PFP, and is not observable for PEN. The beginning of
another weak band in the range of $3.25-3.5$~eV for PEN corresponds to the
tail of the strong $\beta$-band in solution~\cite{Birks}.

In both the intensity distributions of the first absorption band of PEN and
PFP a strong split feature is observed at low energy in contrast to the
solution spectra.  For the second absorption band of PFP the intensity maximum
shifts from the first to the second vibronic excitation.

\section{Discussion}
PEN thin films grown at 30$\, ^\circ$C on Si substrates have a
$d(001)$-spacing of 15.4~\AA, for which the molecular arrangement within the
unit cell is still not established.  Tiago \textit{et
  al.}~\cite{Tiago_2003_PRB_67_115212} calculated optical constants for two
phases of crystalline PEN with a $d(001)$-spacing of 14.1~\AA~and 14.4~\AA,
respectively. The calculated optical spectra of both phases show significant
differences to each other with regard to absolute values and peak positions, a
quantitative comparison with the 15.4~\AA~'thin film' phase may therefore be
problematic.

Nevertheless, the calculated spectra for both phases show strong excitation at
low energies with transition dipoles along the crystallographic a- and b-axis
(overlap of $\pi$-bonds) with an excitonic character and weak excitation with
a transition dipole along the crystallographic c-axis.

This is also valid for the 'thin film phase' we investigate in this paper; the
a- and b-axis of the thin film crystallites coincide with the measured
in-plane component while the out-of-plane component corresponds nearly to the
crystallographic c-axis.  In agreement with calculations from
Ref.~\onlinecite{Tiago_2003_PRB_67_115212} the measured out-of-plane component
exhibits a very weak transition dipole, whereas the in-plane component shows a
strong first absorption feature. The pronounced red-shift of this in-plane
absorption band in the thin film spectra relative to the monomer can thus be
attributed to the generation of excitons.

Due to the observed structural and optical similarities of PEN and PFP one may
speculate that for PFP the same excitonic behavior is present. Then the change
in the first absorption band can also be explained by strong exciton
generation. The behavior of the second band, with a transition dipole nearly
perpendicular to the first band and which corresponds to the $S_0\rightarrow
S_3$ transition of the monomer is different, its excitation energies are
nearly the same as those for the monomer apart from a small solvent shift to
the blue (40~meV).

\section{Summary and Conclusion}
In conclusion, we have provided a comparison of PEN and PFP thin films.
Structurally, the two materials follow similar motifs, although quantitatively
slightly different with regard to the $d(001)$ lattice spacing and the
morphology.

The spectra for PEN and PFP thin films were presented and compared to each
other.  A comparison of our results for the PEN 'thin film phase' with those
obtained by Faltermeier \textit{et al.}~\cite{Faltermeier_2006_PRB_74_125416}
using an effective isotropic fit shows quantitative agreement regarding peak
positions, but differ in relative and absolute intensities. We used an
alternative approach by modeling the 'thin film phase' of PEN anisotropically,
with the optic axis normal to the sample surface (out-of-plane component) and
isotropic properties in the sample plane (in-plane component). With this
approach we can determine the directions of the transition dipoles for each
absorption band, which for the first band in PEN is approximately
perpendicular to the surface normal.  Significant differences were detected
for the spectrum of the thin film compared to those of the monomer. These
changes can be explained by strong exciton generation calculated by Tiago
\textit{et al}.

Two absorption bands were found in PFP thin films, the first has a transition
dipole approximately perpendicular to the surface normal and shows a strong
redshift in respect to the spectrum of the monomer, which is very similar to
the behavior of the first absorption band of PEN.  The transition dipole of
the second band however lies nearly parallel to the surface normal. For this
band the molecular states remain intact apart from a slight blue shift.

Given these results one observes that coupling between molecules in the
condensed phase for PEN and PFP leads to strong effects on the optical
spectra. This may be due to exciton generation in the plane where strong
$\pi$-overlap occurs, but the molecular states with a transition dipole
perpendicular to this plane remain nearly unchanged.


\section{Acknowledgments}
Financial support by the EPSRC and the University of T\"ubingen is gratefully
acknowledged.  We thank Kanto Denka Kogyo Co., Ltd. for providing the PFP.

\end{document}